# Quantifying and mapping covalent bond scission during elastomer fracture

Juliette Slootman[1], Victoria Waltz[1], C. Joshua Yeh[1], Christoph Baumann[2,3], Robert Göstl[2], Jean Comtet*[,1] and Costantino Creton*[,1]

[1]*Laboratory of Soft Matter Science and Engineering, ESPCI Paris, PSL University, CNRS, Sorbonne Université, 75005 Paris, France*
[2]*DWI – Leibniz Institute for Interactive Materials, Forckenbeckstr. 50, 52056 Aachen, Germany*
[3]*Institute of Technical and Macromolecular Chemistry, RWTH Aachen University, Worringerweg 1, 52074 Aachen, Germany*

*costantino.creton@espci.psl.eu, jean.comtet@espci.psl.eu

**Abstract**

**Many new soft but tough rubbery materials have been recently discovered[1-4] and new applications such as flexible prosthetics[5], stretchable electrodes[6] or soft robotics[7] continuously emerge. Yet, a credible multi-scale quantitative picture of damage and fracture of these materials has still not emerged, due to our fundamental inability to disentangle the irreversible scission of chemical bonds along the fracture path from dissipation by internal molecular friction[8]. Here, by coupling new fluorogenic mechanochemistry[9] with quantitative confocal microscopy mapping, we uncover how many and where covalent bonds are broken as an elastomer fractures. Our measurements reveal that bond scission near the crack plane can be delocalized over up to hundreds of micrometers and increase by a factor of 100 depending on temperature and stretch rate, pointing to an intricated coupling between strain rate dependent viscous dissipation and strain dependent irreversible network scission. These findings paint an entirely novel picture of fracture in soft materials, where energy dissipated by covalent bond scission accounts for a much larger fraction of the total fracture energy than previously believed. Our results pioneer the sensitive, quantitative and spatially-resolved detection of bond scission to assess material damage in a variety of soft materials and their applications.**

The failure and fracture of soft materials is an inherently multiscale process: the propagation of a macroscopic crack in the material couples molecular covalent bond scission processes at the crack tip, with deformation and energy dissipation in the bulk [10,11]. Elastomers, a representative class of soft materials, are networks of connected flexible polymer chains, which do not display a well-defined localized yielding behavior, such as metals, ceramics, or polymer glasses. When a crack propagates in an elastomer, it is thus impossible to detect when and where bonds break with conventional methods and the 'process zone' (the mechanically damaged region) is treated for lack of information as an energy sink [12-14] or a cohesive zone of zero thickness[15,16]. Within the field of mechanochemistry, synthetic chemists have recently developed new molecules that can respond optically to forces and bond scission processes when suitably incorporated in polymer networks [9,17,18]. These mechanosensitive molecules have been incorporated into elastomers and have given new insights into molecular fracture processes [19,20]. However, the quantification of bond scission during failure has been elusive due to the lack of suitable mechanophores and of an accompanying robust analytical method [21]. Göstl, Sijbesma, and co-workers have recently reported a fluorogenic mechanophore based on a Diels-Alder (DA) adduct of π-extended anthracene [9,21], that fluoresces stably and sensitively upon force-induced bond scission making it an ideal candidate for quantification of local chain damage. We incorporated this mechanophore as a chain scission-reporting crosslinker into a series of acrylate elastomers prepared by photoinitiated free radical polymerization (Table 1, S1 and SI.1). By fracturing these labeled elastomers at different temperatures and strain rates, we obtain unprecedented quantitative insight into the coupling between molecular bond scission processes at the crack tip and bulk viscoelastic dissipation in these soft materials.

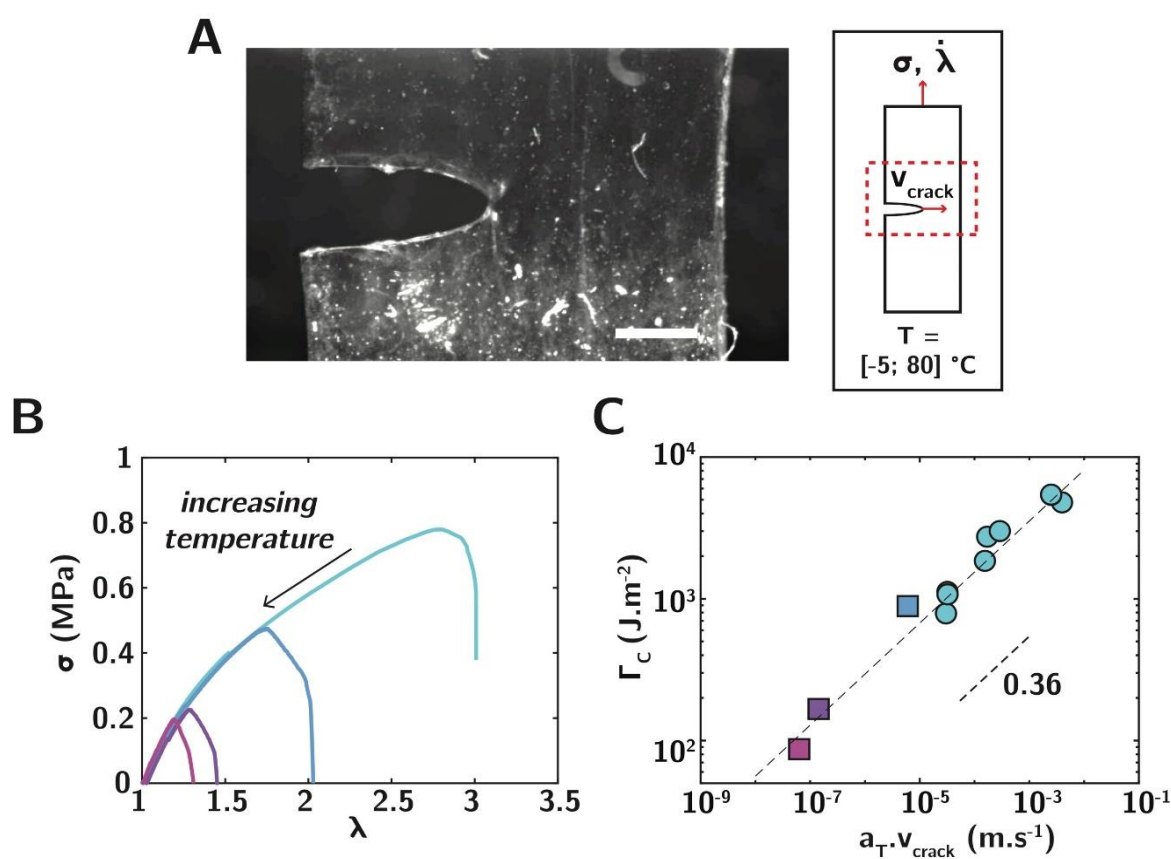

**Figure 1. Macroscopic fracture propagation in elastomers. (A)** *Image of a notched sample during a fracture test. Scale bar is 1 mm. Inset: Geometry of the fracture test, with notched sample in uniaxial extension submitted with a constant stretch rate $\dot{\lambda}$ to an increasing stress $\sigma$ until a crack propagates at speed $v_{crack}$.* **(B)** *Stress strain curves for notched PMA-DA-0.4 elastomer samples at temperature $T = 25, 40, 60$ and $80$ °C (from light blue to purple)* **(C)** *Variation of the fracture energy $\Gamma_c$ as a function of rescaled crack velocity $a_T \cdot v_{crack}$. Squares correspond to samples fractured at different temperature and constant stretch rate $\dot{\lambda} = 3.10^{-3}$ s$^{-1}$ and circles to samples fractured at 25 °C and varying stretch rates $\dot{\lambda} = [3.10^{-4}; 3.10^{-3}; 3.10^{-2}]$ s$^{-1}$.*

|  | Modulus $E$ | Glass transition Temperature $T_g$ | Crosslink density $\nu_x$ | C-C bonds per strand $N_x$ | Strand areal density $\Sigma_{LT}$ |
|---|---|---|---|---|---|
| PMA-DA-0.4 | 1.15 MPa | 18 °C | 4.6 $10^{25}$ m$^{-3}$ | 370 | 1.9 $10^{17}$ m$^{-2}$ |
| PMA-DA-0.2 | 1 MPa | 18 °C | 2.8 $10^{25}$ m$^{-3}$ | 620 | 1.5 $10^{17}$ m$^{-2}$ |
| PEA-DA-0.5 | 1 MPa | -18 °C | 4.2 $10^{25}$ m$^{-3}$ | 400 | 1.8 $10^{17}$ m$^{-2}$ |

**Table 1. Sample and material parameters.** *PMA-DA-0.4, PMA-DA-0.2 and PEA-DA-0.5 are synthetized with 0.02 mol% of DA mechanophore and respectively 0.43, 0.22 and 0.5 mol% of total crosslinker. See SI.1 for details on material synthesis. The cross-link density $\nu_x$ is extracted from fits of the stress-strain curve (see SI.3)*

**Fracture propagation in elastomers.**

We propagate cracks by stretching single-edged notched samples of elastomeric networks in uniaxial extension at different stretch rates $\dot{\lambda}$ and temperatures $T$ (Fig. 1A, see SI. *S2.*). The networks are synthesized from ethyl acrylate (EA) or methyl acrylate (MA) monomers and 1,4-butanediol diacrylate (BDA) as non-mechanoresponsive crosslinker, and are labelled with additional 0.02 mol% (relative to monomer) of the mechanophore diacrylate crosslinker (see Table 1, S1 and synthesis details in *SI.1*). The total crosslink density is of the order of $\nu_x \sim 10^{25}$ m$^{-3}$. These mechanical measurements (Fig. 1B) are used to extract the macroscopic fracture energy $\Gamma_c$ [J·m$^{-2}$] (energy necessary to propagate a unit area of crack), using a fracture mechanics method [22], as well as the crack propagation speed $v_{\text{crack}}$ (see *SI.2*). For the poly(methyl acrylate) elastomer (PMA-DA-0.4, see Table 1), we observe a decrease in fracture energy for increasing temperature, and for decreasing crack propagation speed $v_{\text{crack}}$ (obtained by varying the stretch rate) *(Fig. S6)*, a typical observation in elastomers [23].

As shown in Fig. 1C, following classical time-temperature superposition of the data [13] this macroscopic fracture energy obtained for various temperatures (square) and stretch rates (circles) can be plotted as a sole function of a reduced crack speed $a_T \cdot v_{\text{crack}}$. The factor $a_T$ is a decreasing function of temperature, characterizing viscoelastic dissipation in the sample, and is measured from linear rheology (*Fig. S8*). This overall rescaling leads to a power-law increase in the fracture energy with crack propagation speed (dashed line, Fig. 1C), characterizing the importance of viscoelastic processes during fracture propagation.

For a lack of molecular insights on the actual dissipative processes occurring at the crack tip, this rate dependent behavior has been classically described by the phenomenological expression $\Gamma_c = \Gamma_0 \cdot (1 + f(a_T \cdot v_{\text{crack}}))$, decoupling the rate-independent bond scission processes in the network $\Gamma_0$, occurring strictly in the fracture plane[24], from bulk rate-dependent viscoelastic dissipation (characterized by the function $f(a_T \cdot v_{\text{crack}})$ with $f(v \to 0) = 0$ ), predicted from the linear viscoelastic properties of the material [12,25,26] or simply correlated with mechanical hysteresis [27]. This picture is clearly over-simplified: assuming $\Gamma_0$ to be velocity-independent is in contradiction with the general expectation for rate-dependent materials [28] and quantitative agreement with data requires the introduction of arbitrary length or energy scales[8,29] or predicts viscoelastic dissipation to take place down to typically unphysically small molecular distances at the crack tip [13]. Here, we tackle these inconsistencies in models by quantifying for the first-time molecular bond breakage at the crack tip during fracture propagation in these materials.

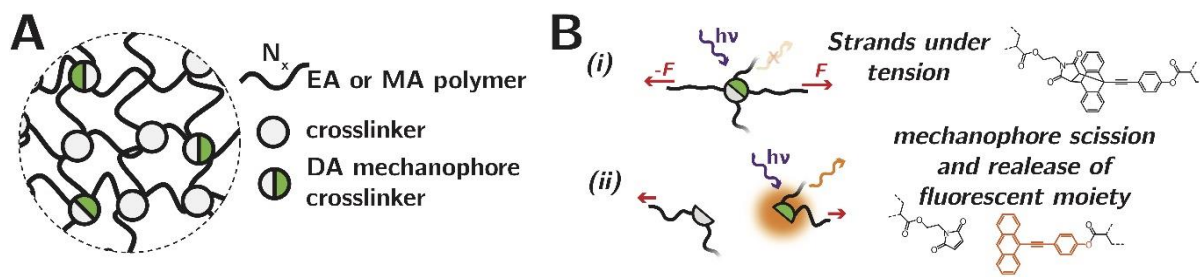

**Figure 2. Strategy for mechanophore incorporation and quantification of the activation.** *(A) Incorporation of mechanophores at crosslink points in the elastomer network. (B) Mechanophore activation reports for strand scission. (i) Non-fluorescent form of the DA mechanophore, connected to a strand under tension. (ii) Irreversible scission of the mechanophore (retro DA reaction), leading to the release of a fluorescent anthracene moiety, reporting strand breakage (orange). Dashed bonds show connectivity to network.*

**Mechanophores quantitatively report strand scission.**

As shown in Fig. 2, by incorporating DA adduct mechanophores as crosslinkers in the network, we can quantify chain scission during elastomer failure (Fig. 2A, typically 5-10% of overall crosslinks are mechanophores). In its native form, the mechanophore is non-fluorescent (Fig. 1B, i). If a sufficient force is applied to the bond, it can undergo cycloelimination (a retro DA reaction), which is irreversible at low temperature [9], leading to the release of a fluorescent π-extended anthracene moiety (Fig. 2B, ii, orange molecule). As previously reported [30,31], the retro DA reaction is greatly accelerated under force with a significantly higher mechanochemical scission rate compared to homolytic C-C bond scission. When connected to the mechanophore, a stressed polymer fragment under extension will thus fail more likely through scission of this mechanochemically weaker bond (Fig. 2B, *ii*), leading to the activation of fluorescence.

Since the mechanophore bond is weaker than the C-C bond, an important question is its ability to quantitatively and reliably report for strand breakage. Given the network heterogeneity and its affine deformation, we hypothesized that the fraction $\phi$ of cleaved chains in the overall material, a characteristic of local damage of the network, is equal to the fraction of cleaved mechanophores. To validate this important hypothesis, we verified that the number of activated mechanophores during mechanical testing varies indeed linearly with the initial fraction of DA adduct used as crosslinker (*Fig. S12*) and that the mechanophore labelled networks have identical mechanical properties as the pristine ones (*Fig. S9*). Although this result may appear surprising, it is due to the fact that even though the mechanophore crosslinker may break at a *weaker force* than a C-C bond it does requires nearly *the same extension* from the strands directly connected to it than would a regular crosslinker.

The extent of strand failure and damage in the material can now be quantified *post-mortem* by measuring the activation of fluorescent mechanophores following crack propagation in various conditions. We used confocal mapping to quantify strand scission in the material normal to the crack surface through the measurement of the local fluorescence intensity due to mechanophore activation (Fig. 3B). Confocal mapping reduces out-of-focus light and allows the measurement of intensity in local volumes x×z×y ("voxels") of typically 1.8×1.8×12 µm$^3$ inside of the material *(see SI. 4)*

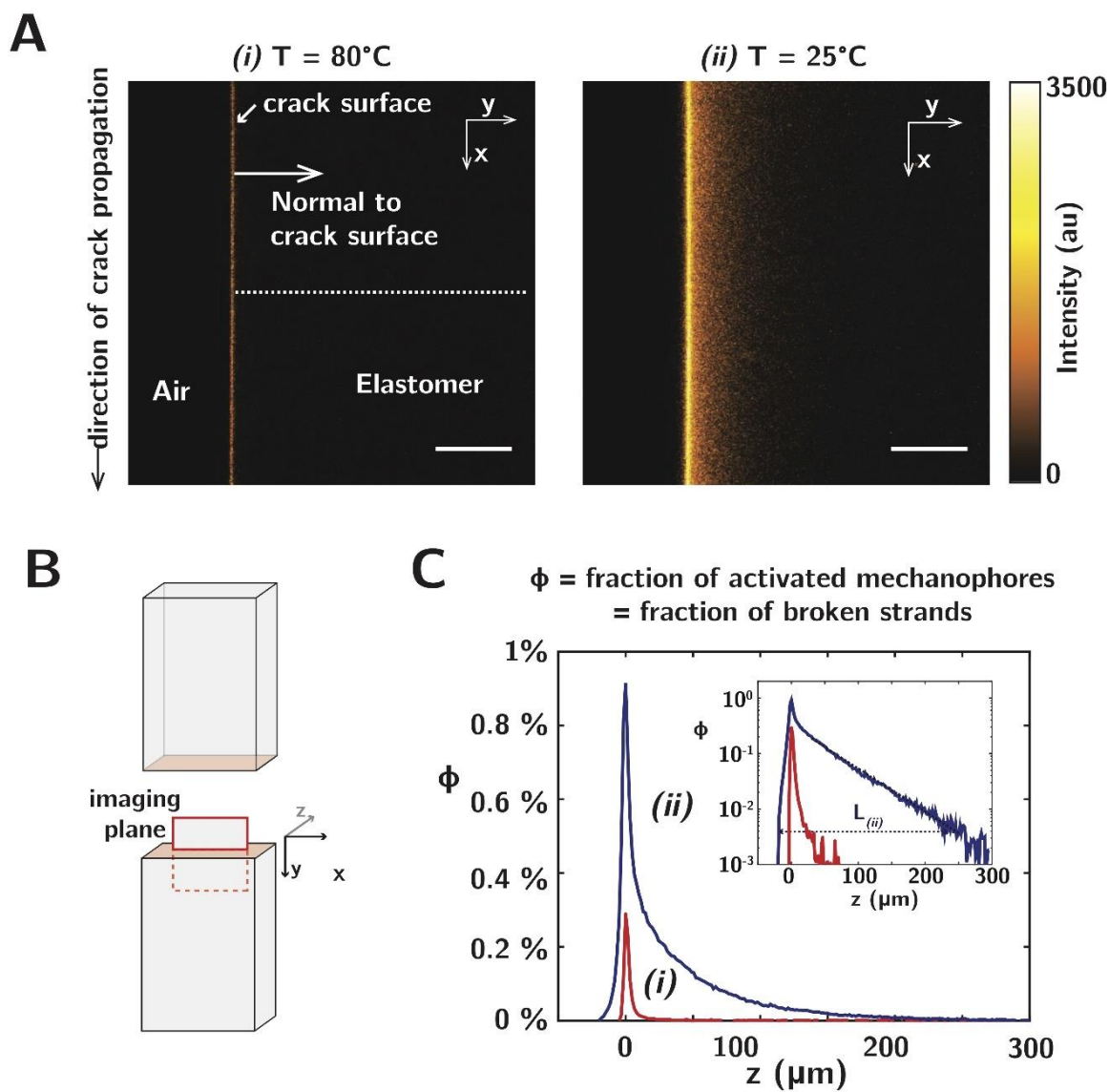

**Figure 3. Post-mortem damage quantification through confocal imaging.** *(A) Post-mortem image of fluorescence activation in a poly(methyl acrylate) sample (PMA-DA-0.4, Table 1) measured by confocal fluorescence microscopy. Samples were fractured at $\dot{\lambda} = 3.10^{-3}$ $s^{-1}$, respectively in conditions of (i) low and (ii) high viscoelasticity, at (i) $T = 80\ °C$ and (ii) $T = 25\ °C$. Pixel size is 1.63 µm. Scale bar is 100 µm. The direction of crack propagation is along the x-direction (vertical). (B) Schematic of the confocal imaging plane (in red), perpendicular to the crack surface (shown in orange). The direction of crack propagation is along the x-direction, the stretch direction along y and sample thickness along z. (C) Average spatial damage profile ϕ for conditions (i) and (ii), with damage ϕ defined as the fraction of broken strands in the material, equal to the fraction of activated mechanophores. Inset shows the profiles in lin-log scale, with $L_{(ii)}$ characterizing the spatial extension of damage for condition (ii).*

Fig. 3A shows 500×500 µm² maps of the fluorescence intensity in planes normal to the crack surface, in two PMA-DA-0.4 samples fractured at stretch rates $\dot{\lambda}$ = 3.10⁻³ s⁻¹ and temperature *(i) $T = 80\ °C$* and *(ii) $T = 25\ °C$*, conditions where *(i)* low and *(ii)* high bulk viscoelastic dissipation is active (Fig. S8D). We observe in these 2D maps a maximum in fluorescence intensity at the crack surface and an intensity profile relatively invariant along the directions of crack propagation (vertical x direction in figure 3A). Remarkably, large differences in the fluorescence profile are observed when comparing these two fracture conditions. For the first sample, fractured at $T = 80$ °C, bulk viscoelastic dissipation is low, and fluorescence activation (and hence network scission) is spatially confined to a region of a few micrometers wide at the crack surface. When the sample is fractured at $T = 25$ °C, where bulk viscoelastic dissipation is much higher, we observe both an increase in intensity revealing a larger local density of broken bonds, as well as a much larger spatial extension of the damage, with strand scission progressively decreasing toward the bulk of the material over hundred micrometers. As similar damage maps are observed along the sample thickness (z-direction, Fig. 3B), we restrict all further quantitative analysis to a constant depth of 100 µm in our samples (see *SI. S4*). Note that a significant thermal contribution to the retro DA fluorescence activation can be ruled out, as this would lead to increasing fluorescence with increasing temperature (*SI. S4*).

For quantification, the local fluorescence intensity is then compared to calibration samples with known amounts of 9-((4-anisyl)ethynyl)anthracene reference fluorophore[9] (equivalent to the activated anthracene, Fig. 2B, *ii*) (*SI. S4*). As shown in Fig. 3C, we extract from these raw confocal images the spatial profile of the fraction $\phi$ of activated mechanophores, equivalent to the fraction of broken strands (Fig. 3C, averaged spatial profile shown respectively in red *(i)* and blue *(ii)*). As already shown in Fig. 3A, the damage profile varies strongly with fracture temperature. We define a damage length $L$, characterizing the spatial extension of strand scission in the material down to the detection threshold concentration of $4.10^{-3}$ %, and equal respectively to $L_{(i)} = 25$ µm and $L_{(ii)} = 250$ µm at 80 °C and 25 °C (Fig. 3C, inset). For sample *ii*, the fraction of broken bonds $\phi$ in the bulk material, at 50 µm from the crack surface, is of the order of 0.1%, corresponding to $10^{22}$-$10^{23}$ broken strands per m³ (one strand every ~ 20-40 nm). As shown in the inset, this damage profile $\phi(z)$ decays here approximately exponentially in the bulk of the material.

To quantify further the extent of damage in the network in each condition, we compute the density $\Sigma$ of cleaved strands per unit area of crack surface created. This quantity $\Sigma = 2 \nu_x \int \phi(z)dz$ (with $\nu_x$ the volume density of cross-links) is obtained by integrating the damage $\phi(z)$ normal to the crack surface, and is found to be respectively $\Sigma_{(i)} \approx 1.2 \; 10^{18}$ m⁻² and $\Sigma_{(ii)} = 2.2 \; 10^{19}$ m⁻² for respectively 80 °C and 25 °C. The factor 2 in the expression of $\Sigma$ accounts for the fact that each crack surface includes only half of the total damage per unit area. $\Sigma$ can be conveniently normalized by $\Sigma_{LT}$, the *minimum* number of strands that need to be broken for a crack to propagate in the material [24]. $\Sigma_{LT}$ can be estimated as $1/2 \cdot \nu_x \langle R_0^2 \rangle^{1/2}$, with $\nu_x$ the volume density of cross-linking points and $\langle R_0^2 \rangle^{1/2}$ the average distance between crosslinks. Following Gaussian statistics, $\Sigma_{LT}$ can be expressed as a function of material parameters [32] (*SI.3*) and is found to be of the order of $10^{17}$ strands.m⁻² (Table 1). For the two conditions in Fig. 3, we find respectively $\Sigma_{(i)}/\Sigma_{LT} \approx 6$ and $\Sigma_{(ii)}/\Sigma_{LT} \approx 110$, a very large value demonstrating that in this condition, crack propagation in the material involves the failure of many more bonds than a single molecular plane, and extends over distances in the material that are more than 4 orders of magnitude larger than the network mesh size.

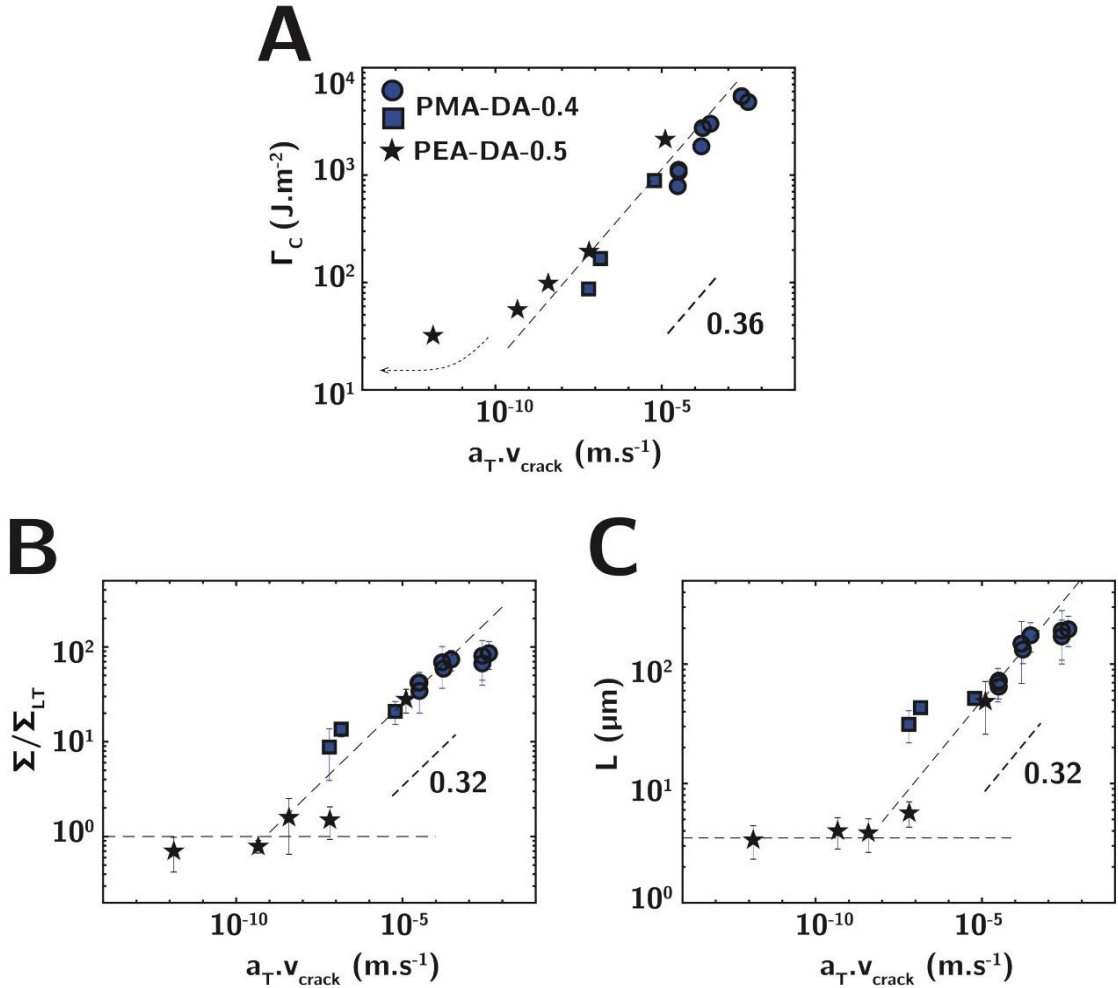

**Figure 4. Coupling of damage with viscoelastic dissipation in the material.** *(A) Fracture energy $\Gamma_c$ as a function of rescaled crack velocity $a_T \cdot v_{crack}$, for the PMA-DA-0.4 (blue squares and circles) and PEA-DA-0.5 samples (black stars). Reference temperature is taken at 25°C for PMA sample (see Fig. S8). At low rescaled crack speed, $\Gamma_c$ becomes less dependent of crack speed for PEA. Blue square corresponds to samples fractured at stretch rate $\dot{\lambda} = 3.10^{-3}$ $s^{-1}$ and different temperature and and blue circles to samples fractured at 25 °C and stretch rates $\dot{\lambda} = [3.10^{-4}; 3.10^{-3}; 3.10^{-2}]$ $s^{-1}$. (B) Normalized areal density of broken strands as a function of rescaled crack velocity. Horizontal line corresponds to the Lake Thomas prediction $\Sigma/\Sigma_{LT} = 1$. (C) Damage length $L$ as a function of rescaled crack velocity. Error bars in Figs. B and C show standard deviation based on 4 local confocal measurements on each fractured sample.*

**Coupling between viscoelastic dissipation and chain damage.**
The strong coupling between damage and viscoelastic dissipation in the sample uncovered in Fig. 3, and the large amount of molecular damage following crack propagation is an unexpected and novel result, never incorporated in any fracture model for lack of experimental insight. Using the same quantification technique, we carried out coupled mechanical measurements and post-mortem damage mapping and quantification, systematically varying the stretch rates and temperature during fracture propagation. In order to increase the range of probed viscoelastic dissipation regimes, we furthermore compared the two PMA and PEA networks of respective glass transition $T_g^{MA} = 18$ °C and $T_g^{EA} = -18$ °C, but similar crosslink density (Table 1). Rescaling between the EA and MA data is obtained by adjusting the viscoelastic shift factor $a(T)$ based on the onset of the glass transition in the storage modulus (*Fig. S8*). As shown in Fig. 4A, we observe for these two materials a power-law increase in the fracture energy with crack propagation speed (dashed lines, Fig. 4A). Due to the low glass transition temperature of the PEA network, we reach for this sample a low-velocity regime for which $\Gamma_c$ becomes less dependent on crack speed, here for $a_T \cdot v_{\text{crack}} < 10^{-10}$ m·s$^{-1}$ (Fig. 4A, dashed arrow). As discussed above, this overall rescaling of $\Gamma_c$ with reduced crack speed $a_T \cdot v_{\text{crack}}$ has been classically accounted for by spatially decoupling rate-independent processes at the crack tip with bulk viscoelastic dissipation.

Our methodology developed in Figs. 2 and 3 allows us to revisit this over-simplified picture, as we can now quantify for each of these experimental conditions the density of strands broken per unit area of crack. To compare the two materials, we plot the normalized damage $\Sigma/\Sigma_{\text{LT}}$ as a function of $a_T \cdot v_{\text{crack}}$. As shown in Fig. 4B, this normalized areal density of broken strands shows a similar trend as that of the fracture energy $\Gamma_c$, with a power-law increase in the density of broken bonds with increasing reduced crack velocity (dashed line in Fig. 4B). This significant increase in covalent bond scission with crack speed contradicts the classical picture of fracture propagation, which assumes strand breakage to be independent or weakly dependent of crack velocity [12,24,33]. At very low crack velocities, we do recover a limiting behavior for which damage appears (within our experimental spatial resolution) indeed solely confined to a molecular plane, as characterized by $\Sigma/\Sigma_{\text{LT}} \approx 1$ (Fig. 4B, horizontal dashed line). We also observe a saturation in damage due to bond scission for the PMA sample in the limit of high crack velocities ($a_T \cdot v_{\text{crack}} > 10^{-3}$ m·s$^{-1}$). Finally, as shown in Fig. 4C, the damage length $L$ characterizing the extension of damage in the material shows a similar trend as the normalized bond breakage $\Sigma/\Sigma_{\text{LT}}$. Viscoelasticity, strand failure and fracture energy appear here strongly coupled.

We rationalize this coupling between viscoelastic dissipation and bond scission in the network as due to an increase of the elasto-adhesive length scale [10] at propagation $\Gamma_c/E$ with increasing reduced crack velocity. In viscoelastic materials, propagating the crack at faster speed leads to more dissipation (the dissipative modulus *E"* increases with strain rate) and requires higher values of the energy release rate *G*, leading to higher strains far from the crack. As discussed in a recent review[10], the elasto-adhesive length at propagation $\Gamma_c/E$ sets the size of the crack tip opening displacement $\delta$ and the scale for the onset of non-linear behaviors at the crack tip. As shown schematically in Figure 5D, this increase of $\delta \sim \Gamma_c/E$ with increased crack speed increases all local strains around the crack tip and hence the local probability of strand scission. Over the range of stretch rates and temperatures probed here, the reduced stretch rate is of the order of $a_T \dot{\lambda} \sim 2.10^{-9} - 5.10^{-4}$ s$^{-1}$, for which the elastic component of the modulus $E' \approx 1$ MPa only varies little with strain rate (Figure S8). This approximately constant modulus rationalizes the similar scaling of $\Gamma_c$, $\Gamma_c/E$ and $\Sigma$ with the reduced crack speed $a_T \cdot v_{\text{crack}}$ (Fig. 4A-B), while the saturation of damage for the largest crack velocity could be due to the onset of stiffening of the material (increase in *E*) for the largest stretch rates (Fig. S8). In essence, this coupling means that a small increase in viscoelastic dissipation per unit volume far from the tip (but over a large volume) can cause a commensurable amount of dissipated energy close to the crack tip by bond scission and by the dissipative processes associated with bond scission.

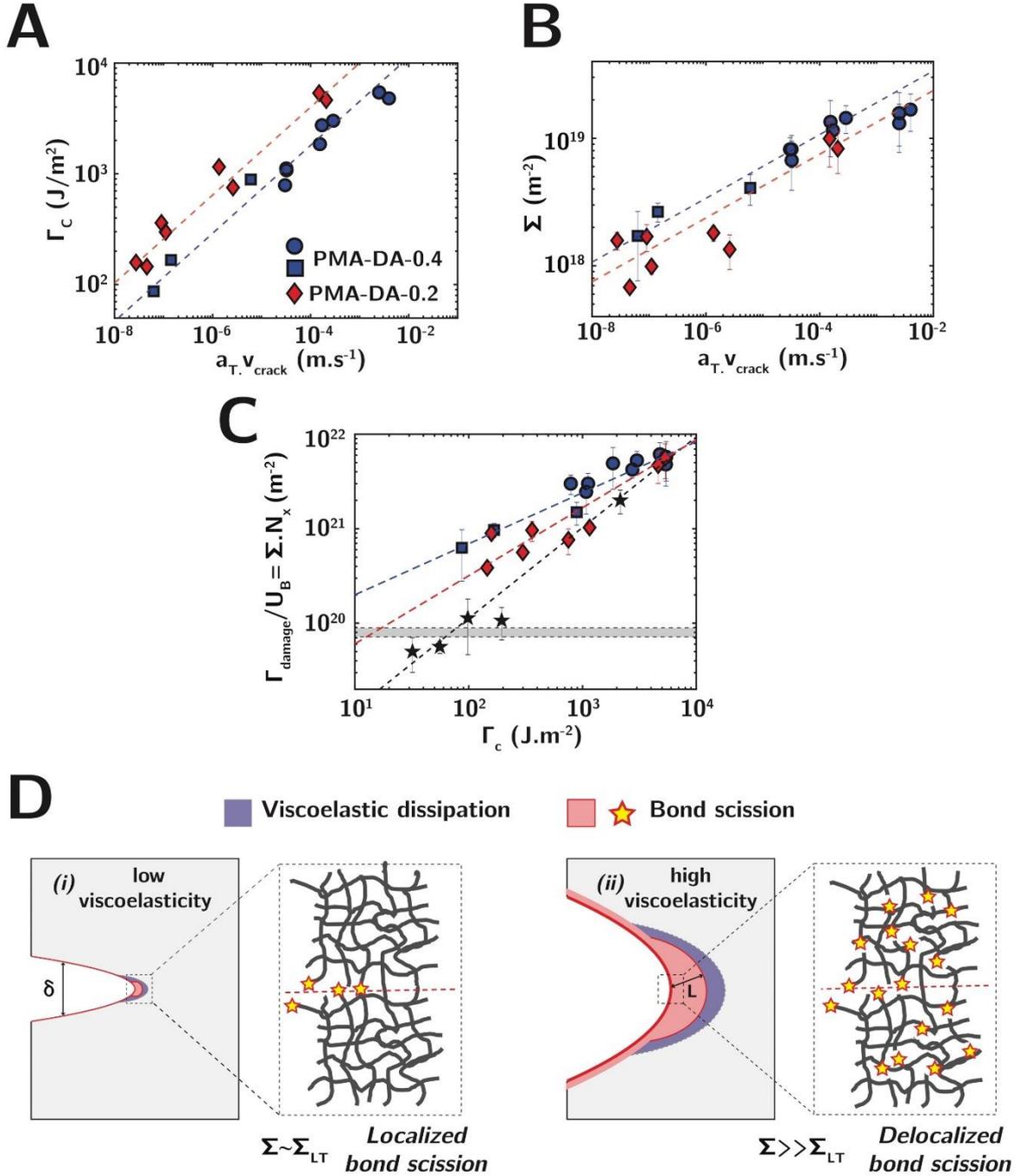

**Fig 5: Effect of molecular structure and contribution of bond scission to the fracture energy.** *(A) Fracture energy as a function of rescaled velocity for two PMA samples with distinct crosslink densities (see Table 1). Dashed lines show power-law fit. (B) Absolute number of cleaved strands per unit area as a function of rescaled velocities for the two samples. Dashed lines show power-law fits. (C) Rescaled fracture energy due to bond scission $\Gamma_{damage}/U_B = \Sigma \cdot N_x$ as a function of the total fracture energy $\Gamma_C$. Dashed lines are power-law fit with power $\beta = 0.95, 0.72,$ and $0.54$ respectively for PEA-DA-0.5, PMA-DA-0.4 and PMA-DA-0.2. The grey area represents $\frac{\Gamma_0}{U_B} = \Sigma_{LT} \cdot N_x$, the Lake-Tomas threshold for the three materials. Error bars in Figs. B and C show standard deviation based on 4 local confocal measurements on each fractured sample. (D) Schematic coupling between viscoelasticity (blue domain) and strand breakage (red domain) at the crack tip. Zoom-in region shows the occurrence of bond scission*

*(yellow stars) in the elastomer network. Bond scission and viscoelastic dissipation are strongly coupled, with joint increase in bond scission and viscoelastic dissipation between the low viscoelasticity (i) and large viscoelasticity regimes (ii). δ represents the crack tip opening displacement and L the characteristic spatial extension of bond scission at the crack tip.*

**Effect of molecular structure.**
Our methodology allows us to further investigate the effects of the molecular architecture of the material, such as the crosslink density, on bond scission at the crack tip. As shown in Fig. 5A, when plotting the fracture energy as a function of rescaled velocity for the two PMA samples with different crosslink densities, we observe that $\Gamma_c$ is larger for the sample with the lowest crosslink density $\nu_x$ (Fig. 5A, comparing red and blue points), but follows the same power law with $a_T \cdot v_{\text{crack}}$. Intriguingly, as shown in Fig. 5B, the density $\Sigma$ of strands cleaved during crack propagation shows a similar scaling, the broken strand density is actually slightly smaller for the PMA-DA-0.2 sample, with the smaller cross-link density and the larger $\Gamma_c$ (Fig. 5B, comparing red and blue points).

We can interpret these trends in the molecular framework set by Lake and Thomas [24]. The classical Lake-Thomas model makes two important claims. First, in the absence of viscoelastic dissipation (in threshold conditions) the fracture energy $\Gamma_c$ is expected to be proportional to the areal density of strands crossing the interface $\Sigma_{\text{LT}}$. Second each broken strand dissipates an energy $N_x \cdot U_B$, with $N_x$ the number of backbone bonds in the strand and $U_B$ the rupture energy of a single bond. While these claims are reasonable in threshold conditions (in the absence of viscoelastic dissipation), where they have been checked experimentally [34-36], there are no obvious reasons for the extension of this coupling between strand scission and fracture energy, when additional viscoelastic dissipative processes are at stake in the material. Our measurements nevertheless demonstrate that a looser network leads to more total dissipated energy for fewer broken strands not only in threshold conditions but also for a wider range of crack propagation speeds.

**Contribution of bond scission to the fracture energy.**
Since a significant level of bond scission occurred in our material during crack propagation, an important question relates to understanding the relative contribution of viscoelastic dissipation and bond scission to the measured fracture energy. Extrapolating the argument of Lake Thomas on strand failure to strands in the bulk, we approximate the energy dissipated per unit area due to bond breakage as $\Gamma_{\text{damage}} = \Sigma \cdot U_{\text{strand}} = \Sigma \cdot N_x \cdot U_B$. The original Lake and Thomas model proposed a value $U_B$ of 350 kJ·mol$^{-1}$ or 3.6 eV for a carbon-carbon bond [24]. A recent analysis of the statistical aspect associated with strand failure based on single molecule stretching experiments provides the value of 60 kJ·mol$^{-1}$ or 0.6 eV/bond as a sounder estimate[37] for the average energy lost by each C-C bond when the polymer strand breaks. Regardless of the exact value associated with this bond scission energy, we plot in Fig. 5C the quantity $\Gamma_{\text{damage}}/U_b = N_x \Sigma$ (directly proportional to the energy dissipated by bond scission) as a function of the macroscopic fracture energy $\Gamma_c$ of the material.

Whereas the energy $\Gamma_{\text{damage}}$ dissipated due to bond scission is classically assumed to be constant and proportional to $\Sigma_{\text{LT}} \cdot N_x$ (horizontal grey domain, Fig. 5C) we find in Fig. 5C a strong correlation between $\Gamma_{\text{damage}}/U_b$ and $\Gamma_c$ (dashed lines are power-law fits). Bond scission thus contributes to the total fracture energy to a much larger extent than previously thought. As schematically represented in Fig. 5D, this energy dissipated by strand scission (red domain) increases roughly in a similar way as the energy dissipated by molecular friction (blue domain) and can reach values of 100 times the Lake Thomas $\Gamma_0$ threshold, i.e. for the PMA-DA-0.4 network, more than 1 kJ/m², dissipated over a volume with dimensions of the order of 100 µm near the crack tip. This new experimental insight highlights the so far neglected influence of molecular bond scission at the crack tip on fracture energies and explains why, as already pointed out by Gent [13], purely linear viscoelastic theories typically require viscoelastic dissipation to take place down to unphysically small molecular distances to the crack tip to account for experimentally measured fracture energy.

Our measurements allow us to further compare specifically the behavior of the different networks (Fig. 5C). We approximate the coupling between bond scission and fracture energies as power-laws, with $\Gamma_{\text{damage}} \sim \Gamma_c^\beta$ with $\beta = 0.95, 0.72$ and $0.54$ respectively, for the three networks. The fraction of energy dissipated through molecular damage and viscoelasticity can be expressed respectively as $\frac{\Gamma_{\text{damage}}}{\Gamma_c} \sim \Gamma_c^{-(1-\beta)}$ and $\frac{\Gamma_{\text{visco}}}{\Gamma_c} \sim \frac{\Gamma_c - \Gamma_{\text{damage}}}{\Gamma_c} \sim 1 - \Gamma_c^{-(1-\beta)}$. The scaling coefficient $\beta \lesssim 1$ for all three networks suggests a slow and progressive transition from a strand scission dominated regime, to a viscoelasticity dominated regime, without any simple decoupling between $\Gamma_{\text{damage}}$ and $\Gamma_{\text{visco}}$.

**Conclusion**
The labelling of elastomeric networks with fluorogenic mechanophores that become fluorescent upon scission give unprecedented insights in the bond scission processes occurring at the crack tip as the materials breaks. Our measurements unveil that, contrary to previous belief, bond scission in these simple networks can extend over more than 100 µm from the crack plane and is strongly dependent on the bulk viscoelastic properties of the network. These observations contradict classical models assuming spatial decoupling between strain-rate independent damage processes at the crack tip and bulk viscoelastic dissipation. These new experimental insights suggest instead the occurrence of intrinsically coupled processes between strand scission and viscoelastic relaxation, mediated by an increase in local strains at the crack tip. Bond scission accordingly accounts for a much larger amount of the fracture energy than anticipated, especially in conditions of large viscoelasticity, and is an overlooked key factor to understand and model fracture toughness from molecular structure.

Our methodology and measurements on model networks open far reaching and diverse paths. It can be used to quantitatively reevaluate a number of damage processes as e.g. occurring during soft material long-term failure and reveal previously invisible damage in a nondestructive way. Our study should further guide the engineering and control of dissipative bond scission processes in complex tough soft materials such as engineering elastomers or tough hydrogels and will stimulate the development of new multiscale models and simulations of elastomer fracture, coupling bond scission and viscoelastic behavior.

**Acknowledgements**
We gratefully acknowledge helpful discussions with Prof. Chung-Yuen Hui , Prof. Rint Sijbesma and Prof. Hugh R. Brown. This project has received funding from the *European Research Council (ERC) under the European Union's Horizon 2020 research and innovation program* under grant agreement AdG No 695351. R.G. and C.B. are grateful for support by a Freigeist-Fellowship of the Volkswagen Foundation (No. 92888). Financial support from the European Commission (EUSMI, No. 731019) is acknowledged. Parts of the analytical investigations were performed at the Center for Chemical Polymer Technology CPT, which was supported by the European Commission and the federal state of North Rhine-Westphalia (No. 300088302).

**References**

1   Gong, J. P., Katsuyama, Y., Kurokawa, T. & Osada, Y. Double-network hydrogels with extremely high mechanical strength. *Adv Mater* **15**, 1155-1158 (2003).
2   Matsuda, T., Kawakami, R., Namba, R., Nakajima, T. & Gong, J. P. Mechanoresponsive self-growing hydrogels inspired by muscle training. *Science* **363**, 504-508, doi:10.1126/science.aau9533 (2019).
3   Sun, J.-Y. *et al.* Highly stretchable and tough hydrogels. *Nature* **489**, 133-136 (2012).
4   Filippidi, E. *et al.* Toughening elastomers using mussel-inspired iron-catechol complexes. *Science* **358**, 502-505, doi:10.1126/science.aao0350 (2017).
5   Minev, I. R. *et al.* Electronic dura mater for long-term multimodal neural interfaces. *Science* **347**, 159-163, doi:10.1126/science.1260318 (2015).
6   Keplinger, C. *et al.* Stretchable, Transparent, Ionic Conductors. *Science* **341**, 984-987, doi:10.1126/science.1240228 (2013).
7   Yuk, H. *et al.* Hydraulic hydrogel actuators and robots optically and sonically camouflaged in water. *Nature Communications* **8**, 14230, doi:10.1038/ncomms14230 https://www.nature.com/articles/ncomms14230#supplementary-information (2017).
8   Persson, B. N. J., Albohr, O., Heinrich, G. & Ueba, H. Crack propagation in rubber-like materials. *Journal of Physics-Condensed Matter* **17**, R1071-R1142 (2005).
9   Gostl, R. & Sijbesma, R. P. [small pi]-extended anthracenes as sensitive probes for mechanical stress. *Chemical Science* **7**, 370-375, doi:10.1039/C5SC03297K (2016).
10  Creton, C. & Ciccotti, M. Fracture and Adhesion of Soft Materials: a review. *Rep Prog Phys* **79**, 046601, doi:10.1088/0034-4885/79/4/046601 (2016).
11  Bai, R., Yang, J. & Suo, Z. Fatigue of hydrogels. *European Journal of Mechanics - A/Solids* **74**, 337-370, doi:https://doi.org/10.1016/j.euromechsol.2018.12.001 (2019).
12  Persson, B. N. J. & Brener, E. A. Crack propagation in viscoelastic solids. *Phys Rev E* **71**, 036123 (2005).
13  Gent, A. N. Adhesion and strength of viscoelastic solids. Is there a relationship between adhesion and bulk properties? *Langmuir* **12**, 4492-4496 (1996).
14  Long, R. & Hui, C.-Y. Fracture toughness of hydrogels: measurement and interpretation. *Soft Matter* **12**, 8069-8086, doi:10.1039/C6SM01694D (2016).
15  Schapery, R. A. Theory of Crack Initiation and Growth in Viscoelastic Media .2. Approximate Methods of Analysis. *International Journal of Fracture* **11**, 369-388 (1975).
16  Knauss, W. G. in *Deformation and Fracture of High Polymers* (eds H. Henning Kausch, John A. Hassell, & Robert I. Jaffee) 501-541 (Springer US, 1973).

17  Davis, D. A. *et al.* Force-induced activation of covalent bonds in mechanoresponsive polymeric materials. *Nature* **459**, 68-72 (2009).
18  Chen, Y. *et al.* Mechanically induced chemiluminescence from polymers incorporating a 1,2-dioxetane unit in the main chain. *Nature Chemistry* **4**, 559-562 (2012).
19  Ducrot, E., Chen, Y., Bulters, M., Sijbesma, R. P. & Creton, C. Toughening Elastomers with Sacrificial Bonds and Watching them Break. *Science* **344**, 186-189, doi:10.1126/science.1248494 (2014).
20  Clough, J. M., Creton, C., Craig, S. L. & Sijbesma, R. P. Covalent Bond Scission in the Mullins Effect of a Filled Elastomer: Real-Time Visualization with Mechanoluminescence. *Adv Funct Mater* **26**, 9063-9074, doi:10.1002/adfm.201602490 (2016).
21  Yildiz, D. *et al.* Anti-Stokes Stress Sensing: Mechanochemical Activation of Triplet–Triplet Annihilation Photon Upconversion. *Angewandte Chemie International Edition* **58**, 12919-12923, doi:10.1002/anie.201907436 (2019).
22  Greensmith, H. W. Rupture of rubber. X. The change in stored energy on making a small cut in a test piece held in simple extension. *Journal of Applied Polymer Science* **7**, 993-1002 (1963).
23  Smith, T. L. Ultimate tensile properties of elastomers. I. Characterization by a time and temperature independent failure envelope. *Journal of Polymer Science Part A: General Papers* **1**, 3597-3615, doi:10.1002/pol.1963.100011207 (1963).
24  Lake, G. J. & Thomas, A. G. The strength of highly elastic materials. *Proceedings of the Royal Society of London, series A: Mathematical and Physical Sciences* **A300**, 108-119 (1967).
25  de Gennes, P. G. Soft Adhesives. *Langmuir* **12**, 4497-4500 (1996).
26  Sakumichi, N. & Okumura, K. Exactly solvable model for a velocity jump observed in crack propagation in viscoelastic solids. *Scientific Reports* **7**, doi:10.1038/s41598-017-07214-8 (2017).
27  Grosch, K., Harwood, J. A. C. & Payne, A. R. Breaking Energy of Rubbers. *Nature* **212**, 497-497, doi:10.1038/212497a0 (1966).
28  Knauss, W. G. A review of fracture in viscoelastic materials. *International Journal of Fracture* **196**, 99-146, doi:10.1007/s10704-015-0058-6 (2015).
29  Long, R., Hui, C. Y., Gong, J. P. & Bouchbinder, E. The fracture of highly deformable soft materials: A tale of two length scales. *ArXiv: 2004.03159 [cond-mat.soft]* (2020).
30  Akbulatov, S. & Boulatov, R. Experimental Polymer Mechanochemistry and its Interpretational Frameworks. *Chemphyschem* **18**, 1422-1450, doi:10.1002/cphc.201601354 (2017).
31  Izak-Nau, E., Campagna, D., Baumann, C. & Göstl, R. Polymer mechanochemistry-enabled pericyclic reactions. *Polymer Chemistry* **11**, 2274-2299, doi:10.1039/C9PY01937E (2020).
32  Millereau, P. *et al.* Mechanics of elastomeric molecular composites. *Proceedings of the National Academy of Sciences* **115**, 9110-9115, doi:10.1073/pnas.1807750115 (2018).
33  de Gennes, P. G. Fracture d'un adhésif faiblement réticulé. *Comptes Rendus de l'Académie des Sciences de Paris, Série II* **307**, 1949-1953 (1988).
34  Ahagon, A. & Gent, A. N. Threshold fracture energies for elastomers. *Journal of Polymer Science, Polymer Physics Edition* **13**, 1903-1911 (1975).

35  Gent, A. N. & Tobias, R. H. Threshold tear strength of elastomers. *Journal of Polymer Science: Polymer Physics Edition* **20**, 2051-2058, doi:10.1002/pol.1982.180201107 (1982).

36  Akagi, Y., Sakurai, H., Gong, J. P., Chung, U.-i. & Sakai, T. Fracture energy of polymer gels with controlled network structures. *The Journal of Chemical Physics* **139**, 144905, doi:doi:http://dx.doi.org/10.1063/1.4823834 (2013).

37  Wang, S., Panyukov, S., Rubinstein, M. & Craig, S. L. Quantitative Adjustment to the Molecular Energy Parameter in the Lake–Thomas Theory of Polymer Fracture Energy. *Macromolecules* **52**, 2772-2777, doi:10.1021/acs.macromol.8b02341 (2019).